\documentclass[12pt]{article}

\usepackage{graphicx,xspace,colortbl}

\usepackage{amsmath,amsthm,amsfonts}
\usepackage{fancybox}

\nonstopmode

\newtheorem{theorem}{Theorem}

\newtheorem{proposition}[theorem]{Proposition}

\theoremstyle{definition}
\newtheorem{definition}[theorem]{Definition}

\newtheorem{assumptions}[theorem]{Assumptions}

\newtheorem{remarks}[theorem]{Remarks}

\def\CF{{\cal F}}

\def\CL{{\cal L}}

\def\CP{{\cal P}}

\def\CDS{\mbox{CDS}}

\def\argmax{\mbox{argmax}}

\def\qed{\hfill$\sqcap\kern-8.0pt\hbox{$\sqcup$}$\\}
\def\beq{\begin{eqnarray}}
\def\eeq{\end{eqnarray}}
\def\beqq{\begin{eqnarray*}}
\def\eeqq{\end{eqnarray*}}
\def\beeq{\begin{eqnarray*}}
\def\eeeq{\end{eqnarray*}}

\def\be{\begin{equation}}
\def\ee{\end{equation}}

\newcommand{{\X}}{{\bf X}}
\newcommand{{\x}}{{\bf x}}
\newcommand{{\Z}}{{\bf Z}}
\newcommand{{\z}}{{\bf z}}
\newcommand{{\Y}}{{\bf Y}}
\newcommand{{\y}}{{\bf y}}
\newcommand{{\F}}{{\bf F}}
\newcommand{{\bbeta}}{{\bf \beta}}
\newcommand{{\bsigma}}{{\bf \sigma}}
\newcommand{{\bL}}{{\bf L}}
\newcommand{{\bW}}{{\bf W}}
\newcommand{{\bu}}{{\bf u}}
\newcommand{{\im}}{\mbox{Im}}

\def\CT{{\cal T}}

\title{Statistical Inference for Time-changed Brownian Motion Credit Risk Models}
\author{T. R. Hurd\thanks{This research is supported by the
Natural Sciences and Engineering Research Council of Canada and
MITACS, Mathematics of Information Technology and Complex Systems
Canada.}\ \ and Zhuowei Zhou\\ Dept. of Mathematics and Statistics\\ McMaster
University\\Hamilton ON L8S 4K1\\Canada}
\date{February 11, 2011}

\begin{document}
\maketitle

\begin{abstract}

We consider structural credit modeling in the important special case where the log-leverage ratio of the firm is a time-changed Brownian motion  (TCBM) with the time-change taken to be an independent increasing process.   Following the approach of Black and Cox, one  defines the time of default to be the first passage time for the log-leverage ratio to cross the level zero. Rather than adopt the classical notion of first passage, with its associated numerical challenges, we accept an alternative notion applicable for TCBMs called ``first passage of the second kind''. We demonstrate how statistical inference can be efficiently implemented in this new class of models. This allows us to compare the performance of two versions of TCBMs, the variance gamma (VG) model and the exponential jump model (EXP), to the Black-Cox model. When applied to a 4.5 year long data set of weekly credit default swap (CDS) quotes for Ford Motor Co, the conclusion is that the two TCBM models, with essentially one extra parameter, can significantly  outperform the classic Black-Cox model.  
\end{abstract}



\newpage

\section{Introduction }\label{introduction}

Next to the Merton credit model of 1974 \cite{Merton74}, the Black-Cox (BC) model \cite{BlacCox76}  is perhaps the best known structural credit model. It models the time of a firm's default as the first passage time for the firm's log-leverage process, treated as an arithmetic Brownian motion, to cross zero.  The BC model is conceptually appealing, but its shortcomings, such as the rigidity of credit spread curves, the counterfactual behaviour of the short end of the credit spread curve and the difficulty of computing correlated multifirm defaults, have been amply discussed elsewhere, see e.g.  \cite{Lando04}. Indeed remediation of these different flaws has been the impetus for many of the subsequent developments in credit risk. 

One core mathematical difficulty that has hampered widespread implementation of Black-Cox style first passage models has been the computation of first passage distributions for a richer class of processes one might want to use in modeling the log-leverage process. This difficulty was circumvented in \cite{Hurd09}, enabling us to explore the consequences of using processes that lead to a variety of desirable features: more realistic credit spreads,  the possibility of strong contagion effects, and ``volatility clustering'' effects.   \cite{Hurd09} proposed a structural credit modeling framework where the log-leverage ratio $X_t:=\log (V_t/K(t))$, where $V_t$ denotes the firm asset value process and $K(t)$ is a deterministic default threshold, is a time-changed Brownian motion (TCBM). The time of default is the first passage time of the log-leverage ratio across zero. In that paper, the time change was quite general: our goal in the present paper is to make a thorough investigation of  two simple specifications in which the time change is of L\'evy type that lead to models that incorporate specific desirable characteristics. We focus here on a single company, Ford Motor Co., and show that with careful parameter estimation, TCBM models can do a very good job of explaining the observed dynamics of credit spreads. TCBMs have been used in other credit risk models, for example \cite{Moosbr06}, \cite{FangJonsOostScho10}, \cite{Baxter06a} and \cite{MendCarrLine10}. 

One model we study is an adaptation of the variance gamma (VG) model introduced by \cite{MadaSene90} in the study of equity derivatives, and remaining very popular since then. We will see that this infinite activity pure jump exponential L\'evy model adapts easily to the structural credit context, and that the extra degrees of freedom it allows over and above the rigid structure of  geometric Brownian motion correspond to desirable features of observed credit spread curves. The other model, the exponential (EXP) model, is a variation of the Kou-Wang double exponential jump model \cite{KouWang03}. Like the VG model it is an exponential 
L\'evy model, but now with a finite activity exponential jump distribution. We find that the EXP model performs remarkably similarly to the VG model when fit to our dataset.
 
We apply these two prototypical structural credit models to a dataset, divided into 3 successive 18 month periods, that consists of weekly quotes of credit default swap spreads (CDS) on Ford Motor Company. On each date, seven maturities are quoted: 1, 2, 3, 4, 5, 7, and 10 years.   The main advantages of CDS data over more traditional debt instruments such as coupon bonds are their greater price transparency, greater liquidity, their standardized structure, and the fact that they are usually quoted for more maturities.

Our paper presents a complete and consistent statistical inference methodology applied to this time series of credit data, one that takes full advantage of the fast Fourier transform to speed up the large number of pricing formula evaluations. In our method, the model parameters are taken as constants to be estimated for each 18 month time period: in contrast to ``daily calibration'' methods, only the natural dynamical variables, not the parameters, are allowed to be time varying. 

Section 2 of this paper summarizes the financial case history of Ford Motor Co. over the global credit crisis period. Section 3 reviews the TCBM credit modeling framework introduced in \cite{Hurd09}. There we include the main formulas for default probability distributions, defaultable bond prices and CDS spreads. Each such formula is an explicit Fourier transform representation that will be important for achieving a fast algorithm. Section 4 gives the detailed specification of the two TCBM models under study. Section 5 outlines how numerical integration of the default probability formula can be cast in terms of the fast Fourier transform.   The main theoretical innovation of the paper is the statistical inference method unveiled in section 6. In this section, we argue that the naive measurement equation is problematic due to nonlinearities in the pricing formula, and that an alternative measurement equation is more appropriate. We claim that the resultant inference scheme exhibits more stable and faster performance than the naive method. In Section 7, we outline an approximate numerical scheme that implements the ideal filter of Section 6. The detailed results of the estimation to the Ford dataset are summarized in Section 8. 

 \section{Ford: The Test Dataset}
 We chose to study the credit history of Ford Motor Co. over the 4.5 year period from January 2006 to June 2010. The case history of Ford over this period spanning the global credit crisis represents the story of a major firm and its near default, and is thus full of financial interest. We have also studied the credit data for a variety of other types of firm over this period, and achieved quite similar parameter estimation results. Thus our study of Ford truly exemplifies the capabilities of our modeling and estimation framework. 
 
We divided the period of interest into three nonoverlapping successive 78 week intervals, one immediately prior to the 2007-2008 credit crisis, 
another starting at the outset of the crisis, the third connecting the crisis and its early recovery. We used Ford CDS and US Treasury yield data, taking only Wednesday quotes in order to remove weekday effects.
\begin{enumerate}
  \item Dataset 1 consisted of Wednesday midquote CDS swap spreads $\widehat \CDS_{m,T}$ and their bid-ask spreads $w_{m,T}$ on dates $t_m=m/52, m=1,\dots, M$  for maturities $T\in\CT:=\{1,2,3,4,5,7,10\}$ years for Ford Motor Co., for the $M=78$ consecutive  Wednesdays from January 4th, 2006 to June 27, 2007, made available from Bloomberg. 
   \item Dataset 2 consisted of Wednesday midquote CDS swap spreads $\widehat \CDS_{m,T}$ and their bid-ask spreads $w_{m,T}$  on dates $t_m=m/52, m=M+1,\dots, 2M$  for maturities $T\in\CT:=\{1,2,3,4,5,7,10\}$ years for Ford Motor Co., for the $M=78$ consecutive Wednesdays from July 11, 2007 to December 31, 2008, made available from Bloomberg.
  \item Dataset 3 consisted of Wednesday midquote CDS swap spreads $\widehat \CDS_{m,T}$ and their bid-ask spreads $w_{m,T}$  on dates $t_m=m/52, m=2M+1,\dots, 3M$  for maturities $T\in\CT:=\{1,2,3,4,5,7,10\}$ years for Ford Motor Co., for the $M=78$ consecutive Wednesdays from January 7th, 2009 to June 30, 2010, made available from Bloomberg. 
   \item The US treasury dataset\footnote{Obtained from US Federal Reserve Bank, www.federalreserve.gov/datadownload} consisted of Wednesday yield curves (the ``zero curve'') on dates $t_m=m/52, m=1,\dots, 3M$, for maturities\\  $$T\in\tilde\CT:=\{1m,3m,6m,1y,2y,3y,5y,7y,10y,20y,30y\}$$ for the period January 4th, 2006 to June 30, 2010. 
\end{enumerate}

We note that Ford Motor Company experienced a large number of credit rating changes during this four-and-a-half year period. The history of Standard \& Poors (S \& P) ratings is as follows: BB+ to  
BB- on January 5, 2006; BB- to B+ on June 28, 2006; B+ to B on September 19, 2006; B to B- on July 31, 2008; B- to CCC+ on November 20, 2008. The downgrades 
continued into 2009, with a move from CCC+ to CC on March 4, 2009 and to SD (``structural default'') on April 6, 2009. The latest news was good: on April 13, 2009, 
S \& P raised Ford's rating back to CCC, on November 3, 2009 to B-, and on August 2, 2010 to B+, the highest since the onset of the credit crisis. 

In hindsight we see that Ford never actually defaulted, although it came close. In the following estimation methodology, we consider the non-observation of default as an additional piece of information about the firm. 

\section{The TCBM Credit Setup}
\label{debtmodel}
The time-changed Brownian motion credit framework of \cite{Hurd09} starts with a filtered probability space $(\Omega, \CF, \CF_t, \mathbb{P})$,  which is assumed to support a Brownian motion $W$  and an independent increasing process $G$ where the natural filtration $\CF_t$  contains $\sigma\{G_u, W_v: u\le t, v\le G_t\}$ and satisfies the ``usual conditions''. $\mathbb{P}$ is taken to be the physical probability measure. \begin{assumptions}\label{assumptions}
\begin{enumerate} \item The log-leverage ratio of the firm, $X_t:=\log(V_t/K(t)):=x+\sigma W_{G_t}+\beta\sigma^2 G_t$ is a TCBM with parameters $x>0,\sigma>0$ and $\beta$. The time change $G_t$ is characterized by its Laplace exponent $\psi(u,t):=-\log {\mathbb{E}}[e^{-uG_t}]$ which is assumed to be known {\it explicitly} and has average speed normalized to 1 by the condition 
  \[\lim_{t\to\infty}t^{-1}\partial \psi(0,t)/\partial u=1.\]
  \item The time of default of the firm is the {\it first passage time of the second kind} for the log-leverage ratio to hit zero (see the definition that follows). The recovery at default is modelled by the ``recovery of treasury'' mechanism\footnote{See \cite{Lando04}.} with constant recovery fraction $R\in[0,1)$. 
  \item The family of default-free zero-coupon bond price processes $\{B_t(T),0\le t\le T<\infty\}$ is free of arbitrage and independent of the processes $W$ and $G$.
  \item There is a probability measure $\mathbb{Q}$, equivalent to $\mathbb{P}$ and called the risk-neutral measure, under which all discounted asset price processes are assumed to be martingales. Under $\mathbb{Q}$, the distribution of the time change $G$ is unchanged while the Brownian motion $W$ has constant drift.\footnote{This assumption can be justified by a particular version of the Girsanov theorem. It would be natural to allow the distribution of $G$ to be different under $\mathbb{Q}$, but for simplicity we do not consider this possibility further here.}  We may write $X_t=x+\sigma W^Q_{G_t}+\beta_Q\sigma^2 G_t$ for some constant $\beta_Q$ where $W^Q_u=W_u +\sigma(\beta-\beta_Q) u$ is driftless Brownian motion under $\mathbb{Q}$. \end{enumerate}
\end{assumptions}

 We recall the definitions from \cite{Hurd09} of first passage times for a TCBM $X_t$ starting at a point $X_0=x\ge 0$ to hit zero. 
\begin{definition}
\begin{itemize}
\item  The standard definition of first passage time is the $\CF$ stopping time 
  \be  t^{(1)}=\inf\{t|X_t\le 0\}\ .\ee
  The corresponding stopped TCBM is $X^{(1)}_t=X_{t\wedge t^{(1)}}$. Note that in general $X^{(1)}_{t^{(1)}}\le 0$.
  \item The first passage time of the second kind is the $\CF$ stopping time
  \be t^{(2)}=\inf\{t|G_t\ge t^*\}\ee
  where $t^*=\inf\{t|x+\sigma W_t+\beta\sigma^2 t\le 0\}$. The corresponding stopped TCBM is 
  \be\label{L2def}X^{(2)}_t=x+\sigma W_{G_{t}\wedge t^*}+\beta \sigma^2(G_{t}\wedge t^*)
 \ee
and we note that $X^{(2)}_{t^{(2)}}=0$.
\end{itemize}
\end{definition}

The general relation between $t^{(1)}$ and $t^{(2)}$ is studied in detail in \cite{HurdKuzn09} where it is shown how the probability distribution of $t^{(2)}$ can approximate that of  $t^{(1)}$. For the remainder of this paper, however, we consider $t^{(2)}$ to be the definition of the time of default.

The following proposition\footnote{Equation \eqref{condCHAR} given in \cite{Hurd09} only deals with the case $\beta<0$. The proof of the extension for all $\beta$ is available by contacting the authors.},  proved in \cite{Hurd09},  is the basis for computing credit derivatives in the TCBM modeling framework.
\begin{proposition} \label{mainprop}Suppose the firm's  log-leverage ratio $X_t$ is a TCBM with $\sigma>0$ and that Assumptions 1 hold. \begin{enumerate}
 \item For any $t>0, x\ge 0$ the risk-neutral survival probability $P^{(2)}(t,x):={\mathbb{E}}_ x[{\bf 1}_{\{t^{(2)}>t\}}]$ is given by
  \be \label{P2}
\frac{e^{-\beta x}}{\pi}\int^\infty_{-\infty}\frac{u\sin(ux)}{u^2+\beta^2}e^{-\psi(\sigma^2(u^2+\beta^2)/2,t)} du +(1-e^{-2\beta x}){\bf 1}_{\{\beta>0\}}, \ee
The density for $X_t$ conditioned on no default is
\beq\label{condPDF}
\rho(y;t,x)&:=&\frac{d}{dy}{\mathbb{E}}_ x[{\bf 1}_{\{X_t\le y\}}|t^{(2)}>t]\\
&=&P^{(2)}(t,x)^{-1}{\bf 1}_{\{y>0\}}\frac{e^{\beta(y-x)}}{2\pi }\int_{\mathbb{R}}\left[e^{iu(y-x)}-e^{-iu(y+x)}\right]e^{-\psi(\sigma^2(u^2+\beta^2)/2,t)}du\nonumber
\eeq
The characteristic function for $X_{t}$ conditioned on no default is
\beq\label{condCHAR}
\mathbb{E}_x[e^{ikX_{t}}| t^{(2)}>t]&=&P^{(2)}(t,x)^{-1}\mathbb{E}_x[e^{ikX_{t}}\cdot {\bf 1}_{\{t^{(2)}>t\}}] \\
&&\hspace{-1in}=P^{(2)}(t,x)^{-1}\frac{e^{-\beta x}}{\pi}\int_{\mathbb{R}}\frac{u\sin(ux)}{(\beta+ik)^2+u^2}e^{-\psi(\sigma^2(u^2+\beta^2)/2,t)}du \nonumber \\
&&\hspace{-1in}\quad +\left(e^{ikx}-e^{-ikx-2\beta x}\right)e^{-\psi(\sigma^2(k^2-2i\beta k)/2,t)}\left(\frac{1}{2} {\bf 1}_{\{\beta=0\}} +{\bf 1}_{\{\beta>0\}}\right) \nonumber
\eeq

  \item The time $0$ price $\bar B^{RT}(T)$ of a defaultable zero coupon bond with maturity $T$ and recovery of treasury with a fixed fraction $R$ is  \be
  \bar B^{RT}(T) = B(T)[P^{(2)}(T,x))+R(1-P^{(2)}(T,x))]
  \ee
\item The fair swap rate for a CDS contract with maturity $T=N\Delta t$, with premiums paid in arrears on dates $t_k=k\Delta t, k=1,\dots, N$, and the default payment of $(1-R)$ paid at the end of the period when default occurs, is given by
\be\label{CDSformula}
\CDS(x,T)=\frac{(1-R)\left[\sum_{k=1}^{N-1}[1-P^{(2)}(t_k,x)][B(t_k)-B(t_{k+1})] +B(T)[1-P^{(2)}(T,x)]\right]}{\Delta t\sum_{k=1}^{N}P^{(2)}(t_k,x)B(t_k)}
\ee
\end{enumerate}
\end{proposition}

\begin{remarks}\begin{itemize}
\item We shall be using the above formulas in both measures $\mathbb{P}$ and $\mathbb{Q}$, as appropriate.
  \item 
We observe in \eqref{P2} that the survival and default probabilities are invariant under the following joint rescaling of parameters \be\label{rescaling}
(x,\sigma,\beta)\to(\lambda z,\lambda \sigma,\lambda^{-1}\beta), \mbox { for any } \lambda>0.
\ee
 It follows that all pure credit derivative prices are invariant under this rescaling.
\end{itemize}\end{remarks}
 
 \section{Two TCBM Credit Models}
 
The two credit models we introduce here generalize the standard Black-Cox model that takes $X_t= x+\sigma W_t+\beta\sigma^2 t$. They are chosen to illustrate the flexibility inherent in our modeling approach. Many other specifications of the time change are certainly possible and remain to be studied in more detail. The following models are specified under the measure $\mathbb{P}$: by Assumption \ref{assumptions} they have the same form under the risk-neutral measure $\mathbb{Q}$, but with $\beta$ replaced by $\beta_Q$.

\subsection{The Variance Gamma Model} 

The VG credit model with its parameters $\theta=(\sigma,\beta,b,c,R)$ arises by taking $G$ to be a gamma process with drift defined by the characteristic triple 
$(b, 0, \nu)_0$ with $b\in(0,1)$ and jump measure $\nu(z)= ce^{-z/a}/z, a>0$ on $(0,\infty)$. The Laplace exponent of $G_t$ is
 \be\label{psiVG}\psi^{VG}(u,t):=-\log E[e^{-u G_t}]=t[bu+c\log(1+au)].
  \ee  
and by choosing $a=\frac{1-b}{c}$ the average speed of the time change is $t^{-1}\partial \psi^{VG}(0,t)/\partial u=1$. This model and the next both lead to a 
log-leverage process of L\'evy type, that is, a process with identical independent increments that are infinitely divisible. 
 
\subsection{The Exponential Model} 
The EXP credit model with its parameters $\theta=( \sigma,\beta,b,c,R)$ arises taking by $G$ to be a L\'evy process  with a characteristic triple 
$(b, 0, \nu)_0$ with $b\in(0,1)$ and jump measure $\nu(z)=c e^{-z/a}/a, a>0$ on $(0,\infty)$. The Laplace exponent of $G_t$ is
 \[ \psi^{Exp}(u,t):=-\log E[e^{-u G_t}]=t\left[bu +\frac{acu}{1+au}\right].
  \]  
 and by choosing $a=\frac{1-b}{c}$ the average speed of the time change is $t^{-1}\partial \psi^{VG}(0,t)/\partial u=1$.

\section{Numerical Integration}
Statistical inference in these models requires a large number of evaluations of the integral formula \eqref{P2} that must be done carefully to avoid dangerous errors and excessive costs. To this end, we approximate the integral by a discrete Fourier transform over the lattice
\[\Gamma=\{  u(  k)=-\bar u +k\eta|  k=0, 1, \dots, N-1\}
\]
for appropriate choices of $N,\eta, \bar u:=N\eta/2.$
It is convenient to take $N$ to be a power of $2$ and lattice spacing $\eta$ such that truncation of the  $u$-integrals to $[-\bar u,\bar u]$ and discretization leads to an acceptable error. If we choose initial values $  x_0$ to lie on the reciprocal lattice with spacing $\eta^*=2\pi/N\eta=\pi/\bar u$
\[\Gamma^*=\{  x(\ell)=\ell\eta^*|  \ell=0, 1,\dots, N-1\}
\]
then the approximation is implementable as a fast Fourier transform (FFT):
\begin{eqnarray}\label{eq10}
P^{(2)}(t,x(\ell))&\sim& 
\frac{-i\eta e^{-\beta x(\ell)}}{\pi}\sum^{N-1}_{k=0}\frac{u(k)e^{iu(k) x(\ell)}}{u(k)^2+\beta^2}\exp[-\psi(\sigma^2(u(k)^2+\beta^2)/2,t)]\\
&=&-i (-1)^n\eta e^{-\beta x(\ell)}
\sum^{N-1}_{k=0}\frac{u(k)e^{2\pi ik\ell/N}}{u(k)^2+\beta^2}\exp[-\psi(\sigma^2(u(k)^2+\beta^2)/2,t)]
\end{eqnarray} 
Note that we have used the fact that  $e^{-i N\eta x(\ell)/2}=(-1)^n$ for all $\ell\in \mathbb{Z}$.

The selection of suitable values for $N$ and $\eta$ in the above FFT approximation of \eqref{CDSformula}  is determined via general error bounds proved in \cite{Lee04}. In rough terms, the pure truncation error, defined by taking $\eta\to 0, N\to\infty$ keeping $\bar u=N\eta/2$ fixed, can be made small if the integrand of \eqref{P2} is small and decaying outside the square $[-\bar u,\bar u]$. Similarly, the pure discretization error, defined by taking $\bar u\to\infty, N\to\infty$ while keeping $\eta$ fixed, can be made small if $e^{-|\beta|\bar x}P^{(2)}(\bar x,t)$, or more simply $e^{-|\beta|\bar x}$, is small, where $\bar x:=\pi/\eta$. One expects that the combined truncation and discretization error will be small if $\bar u$ and $\eta=\pi/\bar x$ are each chosen as above. These error bounds for the FFT are more powerful than bounds one finds for generic integration by the trapezoid rule, and constitute one big advantage of the FFT. A second important advantage to the FFT is its $O(N\log N)$ computational efficiency that yields $P^{(2)}$ on a lattice of $x$ values with spacing $\eta^*=2\pi/N\eta=\pi/\bar u$: this aspect will be very useful in estimation. These two advantages are offset by the problem that the FFT computes values for $x$ only on a grid. 

We now discuss choices for $N$ and $\eta$ in our two TCBM models. For $\beta< 0$, the survival function of the VG model \ is
\begin{equation*}
P^{(2)}(0,t,x,\beta)=\frac{e^{-\beta x}}{\pi}\int_{-\infty}^{\infty}\exp[-tb\sigma^2(u^2+\beta^2)/2]\left(1+\frac{a\sigma^2(u^2+\beta^2)}{2}\right)^{-ct}\frac{u\sin{ux}}{u^2+\beta^2}du
\end{equation*}
while for the EXP model
\begin{equation*}
P^{(2)}(0,t,x,\beta)=\frac{e^{-\beta x}}{\pi}\int_{-\infty}^{\infty}\exp\left[-t\left(b\sigma^2(u^2+\beta^2)/2+\frac{ac\sigma^2(u^2+\beta^2)}{2+a\sigma^2(u^2+\beta^2)}\right)\right]\frac{u\sin{ux}}{u^2+\beta^2}du
\end{equation*}
 In both models, the truncation error has an upper bound $\epsilon$ when $\bar u>C|\Phi^{-1}(\epsilon C')|$, where $\Phi^{-1}$ is the inverse normal CDF and $C,C'$ are constants depending on $t$. On the other hand, provided $\beta < 0$, the discretization error will be small (of order $\epsilon$ or smaller) if\\ $N>\frac{\bar u}{2\pi|\beta|}\log\left(\epsilon^{-1}(1+\exp(-2\beta x))\right)$.  Errors for \eqref{condCHAR} can be controlled similarly. 

% In both models, the truncation errors are bounded by the product of a constant $C$ (depending on $t$) and a normal CDF related to $\bar u$. The truncation error has an upper bound $\epsilon$ when $\bar u>\frac{-1}{\sqrt A}\Phi^{-1}(\frac{\epsilon \sqrt A}{2 C\sqrt{2\pi}})$, where $\Phi^{-1}$ is the inverse normal CDF and $A$ is another constant related to t. On the other hand, provided $\beta < 0$, the discretization error will be small (of order $\epsilon$ or smaller) if the number of discretization $N>\frac{\bar u}{2\pi|\beta|}\log(\frac{\epsilon}{1+exp(-2\beta x)})$. Similar error control can be conducted for \eqref{condCHAR}. 
%

\section{The Statistical Method}\label{statmethod}
The primary aim of this exercise is to demonstrate that our two TCBM credit models can be successfully and efficiently implemented to fit market CDS data on a single firm, in this case Ford Motor Company, and to compare these models' performance to the original Black-Cox structural model.  

We were able to reduce the complexity of our models with negligible loss in accuracy by removing what appear to be two ``nuisance parameters''. First, we expect, and it was observed, that parameter estimations were not very sensitive to $\beta$ near $\beta=0$, so we arbitrarily set $\beta=-0.5$. Secondly, we observed insensitivity to the parameter $b$ and a tendency for it to drift slowly to zero under maximum likelihood iteration: since $b=0$ is a singular limit, we set $b=0.2$.  Finally, in view of the rescaling invariance \eqref{rescaling}, and the interpretation of $\sigma$ as the volatility of $X$, without loss of generality we set $\sigma=0.3$ in all models. So specified, the two TCBM models have three free parameters $\Theta=(c,\beta_Q,R)$ as well as three frozen parameters $\sigma=0.3,\beta=-0.5,b=0.2$. The Black-Cox model with its free parameters $\Theta=(\beta_Q,R)$ and frozen parameters $\sigma=0.3,\beta=-0.5$ then nests as the $c=0$ limit inside both the VG and EXP models. 

We summarize the modeling ingredients:
\begin{itemize}
  \item an unobserved Markov process $X_t\in \mathbb{R}^d$;
  \item model parameters $\Theta\in D\subset\mathbb{R}^n$. We augment the vector $\Theta\to(\Theta,\eta)$ to include an additional measurement error parameter $\eta$;
  \item model formulas $F^k(X,\Theta)$ for $k=1,\dots, K$, which in our case are theoretical CDS spreads given by \eqref{CDSformula} for $K=7$ different tenors; 
  \item a dataset consisting of spreads $Y:=\{Y_t\}$
    observed at times $t=1,\dots, M$ where $Y_t=\{Y_t^k\}$ for a term
    structure of $k=1,\dots, K$, plus their associated quoted bid/ask
    spreads $w^k_t$. We use notation $Y_{\le t}:=\{Y_1,\dots,Y_t \}$ and $Y_{< t}:=\{Y_1,\dots,Y_{t-1}\}$ etc.
\end{itemize}

Since we do not attempt to estimate an underlying interest rate model, we treat the US Treasury dataset as giving us exact information about the term structure of interest rates, and hence the discount factors entering into \eqref{CDSformula}. We treat the quoted bid/ask spreads $w^k_t$ as a proxy for measurement error: these will simplify our treatment of the measurement equation. We also treat the non-default status of Ford on each date as an additional observation. 

To complete the framework, an arbitrary Bayesian prior density of $\Theta$ is taken 
  \[ \rho_0(\Theta):=e^{\CL_0(\Theta)}.
  \]
  with support on $D\subset\mathbb{R}^{n+1}$. The statistical method appropriate to a problem like this is thus some variant of a nonlinear Kalman filter, combined with maximum likelihood parameter estimation.

Based on these assumptions, it is rather natural to assume that observed credit spreads provide measurements of the hidden state vector $X$ with independent gaussian errors. Moreover the measurement errors may be taken proportional to the observed  bid/ask spread. Thus a natural  measurement equation is 
\be\label{measure2} Y^k_t= F^k(X_t,\Theta) + \eta  w^k_t \zeta_t^k \ee 
where $\zeta^k_t$ are independent standard gaussian random variables and $\eta$ is constant. In this case the full measurement density of $Y$ would be
\beq\label{m1like} \CF(Y|X, \Theta)=   \prod_{t=1,\dots,M}\prod_{k=1,\dots,K}\left[\frac1{\sqrt{2\pi}\eta w_t^k}\exp\left(-\frac{(Y_t^k-F^k(X_t,\Theta))^2}{2\eta^2(w^k_t)^2}\right)\right]
  \eeq
  
However, we observed an important deficiency that seems to arise in any scheme like this where the measurement equation involves a nonlinear function of an unobserved process $X$. This nonlinearity leads to nonconvexity in the log-likelihood function for $X$, which in turn can destabilize the parameter estimation procedure. For such reasons, we instead follow an alternative scheme that in our problem, and perhaps many others of this type, gives a great improvement in estimation efficiency. It works in our case because the model formula \eqref{CDSformula} for $F^k(x,\Theta)$, although nonlinear in $x$, is monotonic and approximately linear in $x$. We will call our scheme the ``linearized measurement'' scheme and it is justified as follows.

We define $G^k(Y,\Theta)$ to be the solution $x$ of $Y=F^k(x,\Theta)$, and note that $f^k:=\partial_x F^k>0$. 
Then, provided $\eta w^k$ are small enough, we may linearize the $x$ dependence of the measurement equation using the Taylor expansion
\beqq  Y^k-F^k(x)&=&Y^k-F^k(G^k(Y^k)+x-G^k(Y^k))  \\
&\approx&Y^k-F^k(G^k(Y^k))+f^k(G^k(Y^k))(G^k(Y^k)-x) \\
&=&f^k(G^k(Y^k))(G^k(Y^k)-x)
\eeqq
This equation above justifies the following alternative to the measurement equation \eqref{measure2}:
\be
 \tilde X^k_t=X_t +\eta \tilde w^k_t\xi^k_t \ee \\
Now $\xi^k_t, k=1,2,\dots, K, t=1,2,\dots, M$ are iid $N(0,1)$ random variables and the transformed measurements are 
\[\tilde X^k_t=\tilde X^k(Y^k_t,\Theta):=G^k(Y^k_t,\Theta).
\]
Furthermore,
\[ \tilde w^k_t=\tilde w^k_t(\tilde X^k_t,\Theta) =f^k(\tilde X^k_t,\Theta)^{-1} w^k_t. \]
Note that $\tilde X^k_t, k=1,\dots, K$ have the interpretation as independent direct measurements of the unobserved state value $X_t$. 

The full measurement density of $Y$ in our linearized measurement scheme is thus: 
   \beq\label{fulldens}
   \CF(Y|X,\Theta)&:=&\prod_{t=1,\dots,M}f(Y_t|X_t,\Theta)\\
   f(Y_t|X_t,\Theta)&:=&\prod_{k=1,\dots,K}\left[\frac{1}{\sqrt{2\pi}\eta w_t^k}
\exp\left(-\frac{(\tilde X^k_t(Y^k_t,\Theta)-X_t)^2}{2\eta^2\tilde w^k_t(\Theta)^2}\right)\right]  
  \eeq
  where we have recombined denominator factors of $\tilde w^k$ with Jacobian factors $(f^k)^{-1}$. 
  The multiperiod transition density conditioned on nondefault  is 
  \be \CP(X|\Theta,\mbox{no default})=\prod_{{t=2,\dots,M}}p(X_t|X_{t-1},\Theta)
 \label{transitiondensity}\ee
where $p(y|x, \Theta)$ is the one period conditional transition density given by \eqref{condPDF} with $t=\Delta t$. Finally the full joint density for $(X,Y,\Theta)$ is
\be \rho(X,Y,\Theta):= \CF(Y|X,\Theta) \CP(X|\Theta)\rho_0(\Theta) 
\ee
Integration over the hidden state variables $X$ leads to the partial likelihood function, which can be defined through an iteration scheme:
\be \label{partialLH}
\rho(Y, \Theta)=\int f(Y_M|X_M,\Theta) \rho(X_M,Y_{<M},\Theta) dX_M
\ee
where for $t< M$
\be
 \rho(X_{t+1},Y_{\le t},\Theta)=\left\{\begin{array}{ll }
 \int p(X_{t+1}|X_{t},\Theta)f(Y_t|X_{t},\Theta)\rho(X_t,Y_{<t},\Theta) dX_t,\     & t>0\\   \\
 \rho_0(\Theta)     &   t=0
\end{array}\right.\label{inductive}
\ee

  The following summarizes statistical inference within the linearized measurement scheme.\medskip
    
 \noindent{\bf Statistical Inference using the Linearized Measurement Scheme:\ } Let $(\hat Y,w):=\{\hat Y^k_t,w^k_t\}$ be the time series of CDS observations.  
 \begin{enumerate}
  \item Maximum Likelihood Inference: The maximum likelihood parameter estimates $\widehat\Theta$ are the solutions of
  \be\label{ThetaInference}\widehat\Theta=\argmax_{\Theta\in D} \log
  \left(\rho(\hat Y, \Theta)/\rho_0(\Theta)\right)\ee
  where $ \rho(\hat Y, \Theta)$ is given by \eqref{partialLH}.
  The log-likelihood achieved by this solution is
  \[ \widehat \CL:=\log\left(\rho(\hat Y, \widehat \Theta)/\rho_0(\widehat\Theta)\right),\] 
  and the Fisher information matrix is
  \[\widehat{\cal I}:= -\left[\partial^2_{\Theta}\log\left(\rho(\hat Y,
      \widehat \Theta)/\rho_0(\widehat\Theta)\right)\right];\] 
  \item Filtered State Inference: The time series of filtered estimates of the state variables $X_1,\dots, X_M$ are the solutions $\hat X_1,\dots,\hat X_M$ of 
  \be\label{Xinference} \hat X_t=\argmax_{x\in \mathbb{R}_+}\log
  \left(f(\hat Y_t|x,\widehat\Theta)\rho(x,\hat Y_{\le t-1},\widehat\Theta)\right)
  \ee
\end{enumerate}
  
  \section{Approximate Inference}
  
The previous discussion on inference was exact, but computationally
infeasible. Our aim now is to give a natural and simple approximation
scheme that will be effective for the problem at hand. Our scheme is to inductively approximate the likelihood function
$\rho(X_{t+1},Y_{\le t},\Theta)$ defined by \eqref{inductive} by a truncated normal distribution
through matching of the first two moments. The truncation point of $0$ is determined by the no default condition. The rationale is that the non-gaussian nature of the transition density $p$ will have only a small effect when combined with the gaussian measurement density $f$. We expect our approximation to be appropriate for a firm like Ford that spent a substantial period near default. As we discuss at the end of this section, a simpler approximation is available that is applicable to a firm of high credit quality. The more complicated method we now describe is intended to be more robust when applied to firms of a range of credit qualities.

We describe a single step of the inductive computation of $\rho$ given by \eqref{inductive}. We  fix $t$, denote the time $t$ state variable as $x$ and
the time $t+1$ state variable as capital $X$.  The length between $t$
and $t+1$ is denoted as $\Delta t$. We also suppress $Y_{\le t}$ and $\Theta$. 
In this context, we are looking for $\bar{\mu}$ and $\bar{\sigma}$
that satisfy
\be\label{truncnorm}
 \rho(X)\approx\frac{m_0\phi\left(\frac{X-\bar{\mu}}{\bar{\sigma}}\right)}{\Phi\left(\frac{\bar{\mu}}{\bar{\sigma}}\right)} , X>0
\ee 
where
\be m_0=  \int_0^{\infty} f(x)\tilde\rho(x) dx .\ee
Here $\phi$ and $\Phi$ are probability density and cumulative distribution
functions of the standard normal distribution and $\tilde\rho$ is carried over from the previous time step.  The first two moments of the truncated normal distribution are straightforward to derive and are given here for completeness:
\beqq
m_1^{trunc}&=&\bar{\mu}+\bar{\sigma}\lambda(\alpha)\\
m_2^{trunc}&=&\bar{\sigma}^2[1-\delta(\alpha)]+(m_1^{trunc})^2
\eeqq
where $\alpha=-\frac{\bar{\mu}}{\bar{\sigma}}, \lambda(\alpha)=\frac{\phi(\alpha)}{1-\Phi(\alpha)}, \delta(\alpha)=\lambda(\alpha)[\lambda(\alpha)-\alpha]$. Note that the truncated normal distribution has a larger mean and smaller variance than the original normal distribution.
 
Using the Fubini theorem, the first two moments of the distribution $\rho(X)$ are:
\beq\label{approxm}
m_1&=& m_0^{-1} \int_0^{\infty} g_1(x)f(x)\tilde\rho(x) dx \\
m_2&=& m_0^{-1} \int_0^{\infty} g_2(x)f(x)\tilde\rho(x) dx \nonumber
\eeq
Here $g_1(x)$ and $g_2(x)$ are the first and second moments of $X$
with respect to the transition density $p(X|x)$ and are given using \eqref{condCHAR} by
\beq
g_1(x)=\frac{1}{i}\partial _k|_{k=0}\mathbb{E}^P_x[e^{ikX}| t^{(2)}>\Delta t] \\
g_2(x)=-\partial^2_k|_{k=0}\mathbb{E}^P_x[e^{ikX}|t^{(2)}>\Delta t] \nonumber
\eeq

Note that $\tilde \rho(x)$
has a gaussian kernel approximation by induction and the measurement density $f(x)$
is also gaussian. Their product gaussian kernel is then simply a scaled
normal probability density function:
\be
f(x)\tilde\rho(x)=\frac{m_0}{\sqrt{\bar v}\Phi(\bar m/\sqrt{\bar v})}\phi\left(\frac{x-\bar m}{\sqrt{\bar v} }\right)
\ee  
We also notice that the transition density $p(X|x)$ with a short period
$\Delta t$ resembles a Dirac $\delta$ function of $X$ and fitting it to a polynomial would require very
high order to guarantee accuracy in a local domain. In our method, by contrast,  the
moment functions $g_1(x)$ and $g_2(x)$ that appear in the integrals  in \eqref{approxm} are much smoother functions
of $x$ and usually low order polynomials can approximate them quite
accurately in a local domain. Take a normal transition density for
example: $g_1(x)$ is linear in $x$ and $g_2(x)$ is quadratic in
$x$. Their counterparts for time changed Brownian motion conditional on
no default can also be well approximated by low order polynomials in a
local domain. We stress the word ``local'' because  the product gaussian kernel $f\rho$ typically
has a moderate variance $\bar v$ and relatively large mean $\bar m$: therefore the integrals in
Equation \eqref{approxm} are dominated by a local domain $[\bar
m-a\sqrt{\bar v}, \bar  m+a\sqrt{\bar v}]$ with $a$ safely taken to be $4$.  Thus we need to fit $g_1(x)$ and $g_2(x)$ over the interval $[\bar m-a\sqrt{\bar v}, \bar  m+a\sqrt{\bar v}]$ which can be done quite accurately with quartic polynomials: 
\beq
g_1(x)&=&\Sigma_{k=0}^{4}c_{1k}(x-\bar m)^k \\
g_2(x)&=&\Sigma_{k=0}^{4}c_{2k}(x-\bar m)^k. \nonumber
\eeq

Equation \eqref{approxm} is now approximated by
 \beq\label{mapprox2}
m_1&=& \frac{1}{m_0\sqrt{\bar v}\Phi(\bar m/\sqrt{\bar v})} \int_0^{\infty} \Sigma_{k=0}^{4}c_{1k}(x-\bar
m)^k\phi\left(\frac{x-\bar m}{\sqrt{\bar v} }\right) dx \nonumber\\
m_2&=& \frac{1}{m_0\sqrt{\bar v}\Phi(\bar m/\sqrt{\bar v})}\int_0^{\infty} \Sigma_{k=0}^{4}c_{2k}(x-\bar
m)^k\phi\left(\frac{x-\bar m}{\sqrt{\bar v} }\right) dx \nonumber
\eeq
which can be evaluated analytically in terms of the error function.
Matching $m_1$ and $m_2$ with $m_1^{trunc}$ and $m_2^{trunc}$ determines $\bar{\mu}$ and $\bar{\sigma}$ and completes the iteration scheme for \eqref{partialLH}.

 \begin{remarks}\label{rm4}\ \smallskip 
 \begin{itemize}
 \item In our numerical examples, we enlarge the integral domain in
  Equation \eqref{mapprox2} from $\mathbb{R^+}$ to $\mathbb{R}$ if
  $\bar m>4\sqrt{\bar v}$, which leads to a simpler
  implementation. It turns out in our study that this condition is satisfied for all
  sampling periods.
  \item  An alternative moment matching approximation is
   possible which approximates $\rho(X)$ by a regular normal
   distribution, rather than a truncated normal. Then the truncated
   density in Equation \eqref{truncnorm} should be replaced by the regular density
   $\phi\left(\frac{X-\bar{\mu}}{\bar{\sigma}}\right), X\in\mathbb{R}$. Although this
   approximation conflicts with the default barrier, for a firm that
   is far from default this does not introduce a serious numerical error. Moreover, this approximation
   leads to linear gaussian transition density and is thus a  Kalman filter. \end{itemize}
  \end{remarks}
Here we summarize the computation of $\rho(Y_{\le M},\Theta)$ for a fixed value of $\Theta$:  
\begin{enumerate}
  \item Set $\rho_1=\rho_0(\Theta)$;
  \item Compute the measurement density $f(Y_1|X_1)$ (i.e. compute its mean and variance: this step requires efficient use of the FFT to invert the  $\CDS$ spread formula);
\item For $t=1:M-1$
  \begin{enumerate}
   \item Approximate $\rho(X_{t+1},Y_{\le t},\Theta)$ given by
     \eqref{inductive} by a truncated normal density with mean and variance computed by matching moments. For this one uses the exact formula for the first two moments of the conditional transition density \eqref{condPDF}, and the assumed normal form of $f(Y_t|X_t)$ and $\rho(X_{t},Y_{<t},\Theta)$; 
 \item Compute the measurement density $f(Y_{t+1}|X_{t+1})$ (ie. compute its mean and variance, again with efficient use of FFT);
  \item End loop;
\end{enumerate}  \item Finally compute $\rho(Y_{\le M}, \Theta)$  by integrating $X_M$ as in \eqref{partialLH}.
\end{enumerate}

\section{Numerical Implementation}\label{nuresult}

From the considerations described in section \ref{statmethod} we fix $\beta=-0.5, \sigma=0.3, b=0.2$. We choose $\bar u=300$ which controls the truncation error within $10^{-10}$. Depending on $\Theta$, we allowed the size of the FFT lattice, $N$, to vary from $2^{8}$ to $2^{10}$,
keeping the discretization error within $10^{-10}$. We use the Matlab function {\it{fmincon}} to implement the quasi-Newton method to maximize the likelihood function. Since {\it{fmincon}} also calculates the gradient and Hessian of the objective function, we also obtain standard errors of the parameter estimates.

\begin{table}[b] 
\centering\begin{tabular}{||c||c||c|c|c||}
\hline \hline
&&Dataset 1&Dataset 2&Dataset 3\tabularnewline
\hline 
&number of weeks&78&78&78\tabularnewline\hline\hline
&$\hat\sigma$&0.3&0.3&0.3\tabularnewline\hline
&$\hat{b}$&0.2&0.2&0.2\tabularnewline\hline
&$\hat{c}$&1.039(0.060)&0.451(0.034)&1.08(0.11)\tabularnewline\hline 
&$\hat\beta_Q$&-1.50(0.12)&-0.879(0.061)&-1.368(0.066)\tabularnewline
\hline
VG Model&$\hat R$&0.626(0.026)&0.450(0.029)&0.611(0.018)\tabularnewline\hline 
&$\hat\eta$&1.53&0.897&1.797\tabularnewline\hline
&$\hat{x}_{av}$&0.693&0.457&0.480\tabularnewline\hline 
&$\hat{x}_{std}$&0.200&0.239&0.267\tabularnewline
\hline 
&{\rm RMSE} &1.43&0.837&1.792\tabularnewline\hline\hline
&$\hat\sigma$&0.3&0.3&0.3\tabularnewline\hline
&$\hat b$&0.2&0.2&0.2\tabularnewline\hline 
&$\hat{c}$&2.23(0.12)&1.17(0.07)&2.33(0.20)\tabularnewline
\hline
&$\hat\beta_Q$&-1.44(0.12)&-0.780(0.060)&-1.286(0.067)\tabularnewline
\hline
Exponential Model&$\hat R$&0.609(0.028)&0.395(0.033)&0.588(0.022)\tabularnewline\hline
&$\hat\eta$&1.503&0.882&1.775\tabularnewline\hline
&$\hat {x}_{av}$&0.702&0.479&0.486\tabularnewline
\hline
&$\hat {x}_{std}$&0.199&0.242&0.266\tabularnewline
\hline
&{\rm RMSE}&1.41&0.821&1.763\tabularnewline\hline\hline
&$\hat\sigma$&0.3&0.3&0.3\tabularnewline\hline 

&$\hat\beta_Q$&-2.02(0.10)&-1.793(0.067)&-1.78(0.12)\tabularnewline
\hline
&$\hat R$&0.773(0.011)&0.757(0.009)&0.760(0.013)\tabularnewline\hline
Black-Cox Model&$\hat\eta$&2.38&1.29&2.18\tabularnewline\hline
&$\hat {x}_{av}$&0.624&0.406&0.422\tabularnewline
\hline
&$\hat {x}_{std}$&0.187&0.214&0.237\tabularnewline
\hline
&{\rm RMSE}&2.19&1.19&2.14\tabularnewline\hline\hline

\end{tabular}

\caption{Parameter estimation results and related statistics for the VG, EXP and Black-Cox 
models. $\hat X_t$ derived from \eqref{Xinference} provide the estimate of the hidden state variables. 
The numbers in the brackets are standard errors. The estimation uses weekly (Wednesday) 
CDS data from January 4th 2006 to June 30 2010. $\hat {x}_{std}$ is the square root of the annualized
quadratic variation of $\hat X_t$.}\label{calibration1}
\end{table}

\begin{table}[b]
\centering\begin{tabular}{||c|c|c|c||}\hline \hline
&VG&EXP&B-C \tabularnewline\hline 
VG&0&-2.21/-1.41/-2.33 & 5.42/5.10/2.03 \tabularnewline\hline
EXP&2.21/1.41/2.33 &0& 5.46/5.22/2.19\tabularnewline\hline
B-C&-5.42/-5.10/-2.03 &-5.46/-5.22/-2.19 &0\tabularnewline\hline 
\end{tabular}

\caption{Results of the Vuong test for the three models, for dataset 1, dataset 2 and dataset 3. A positive value larger than 1.65 indicates that the row model 
is more accurate than the column model with 95\% confidence level.}\label{modelcomp}
\end{table}

Table \ref{calibration1} summarizes the estimation results for each of the three models, for the three datasets in 2006-2010, using our time
series approximate inference. 
Estimated parameter values are given with standard errors, as well as summary statistics for the resulting filtered time series of $X_t$. 
We also present the root mean square error ({\rm RMSE}) defined as the average error of the CDS spreads quoted in units of the bid/ask spread.
\[
 {\rm {\rm RMSE}}=\sqrt{\frac{1}{M\cdot K}\sum_{t=1}^M\sum_{k=1}^K\frac{\left( F^k(X_t,\Theta)-Y_t^k\right)^2}{(w_t^k)^2}}
\]
 
Overall, the  finite activity EXP model shares quite a few similarities with the infinite activity VG model, both in behavior and performance.  For these two TCBM models, their model parameters are quite similar between dataset 1 and dataset 3 respectively. It is consistent with Ford's history of credit ratings that dataset 3 has lower, more volatile log-leverage ratios and lower recovery rate than dataset 1. We can also see that during the peak of the credit crisis in dataset 2, the estimated parameters show noticeable signs of stress. The mean time change jump size is up by approximately $50\%$, driven mainly by the increased short term default probability. 
The recovery rate is significantly lower. In the very stressed financial environment at that time, a firm's value would be greatly discounted and its capacity to liquidate assets would be limited. On the other hand the risk neutral drift $\beta_Q$ is significantly higher, reflecting a certain positive expectation on the firm. At the peak of the credit crisis, Ford's annualized credit spreads exceeded 100\%. The log-leverage ratios are much suppressed to a level of about 65\% of that of dataset 1. 

By definition, ${\rm RMSE}$ measures the deviation of the observed CDS spreads from the model CDS spreads while $\eta$ measures the deviation of the ``observed'' log-leverage ratios $\tilde X_t$ from the ``true'' log-leverage ratios $X_t$. We can see that ${\rm RMSE}$ and $\eta$ are very close in all cases, which implies that the objective functions based on the naive CDS measurement density \eqref{m1like} and the linearized measurement density \eqref{fulldens} are fundamentally very similar.

\begin{figure}[b]
\centering\includegraphics[%
  scale=.7]{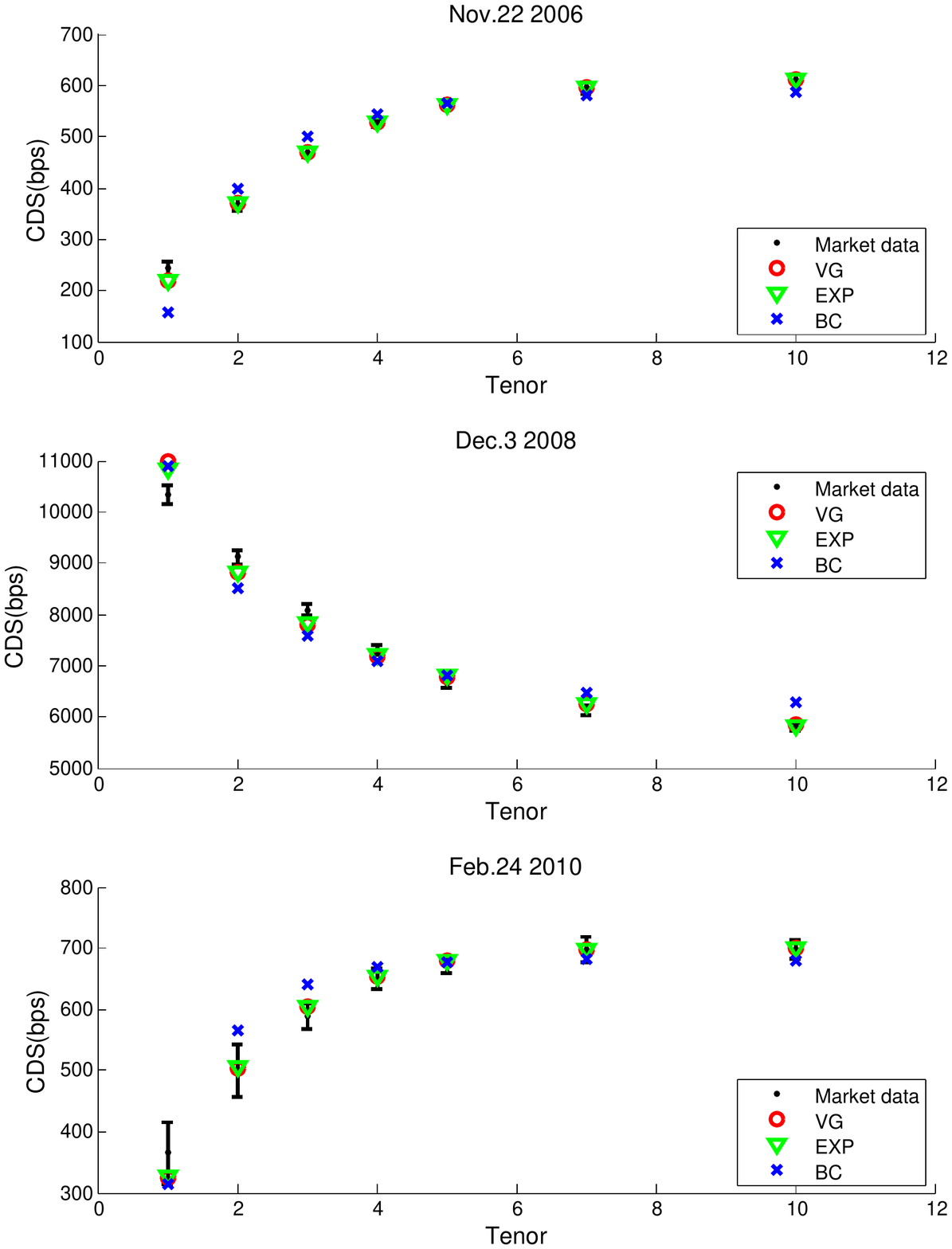}
\caption{{The in-sample fit  of the two TCBM models and Black-Cox model to the observed Ford CDS term structure for November 22, 2006 (top), December 3, 2008 (middle)
and February 24, 2010 (bottom). 
The error bars are centered at the mid-quote and indicate the size of the bid-ask spread.}}
\label{CDScalib0608}

\end{figure}

In terms of ${\rm RMSE}$ and $\eta$, both TCBM models performed much better than the Black-Cox model. The TCBM fitting is typically within two times the bid/ask spread across 3 datasets, while the errors of the Black-Cox model are about $30\%$ higher on average. Figure \ref{CDScalib0608} shows that on three typical days, 
the TCBM models can fit the market CDS term structure curves reasonably well while the Black-Cox model, with its  restrictive hump-shaped term structures, has difficulties for some tenors. 
To fit high short spreads, the log-leverage ratio is forced to unreasonably low levels.
The TCBM models, with only one extra parameter than the Black-Cox model,  generate more flexible shapes, and do a better job of fitting the data.

\begin{figure}[b]
\centering\includegraphics[%
  scale=.7]{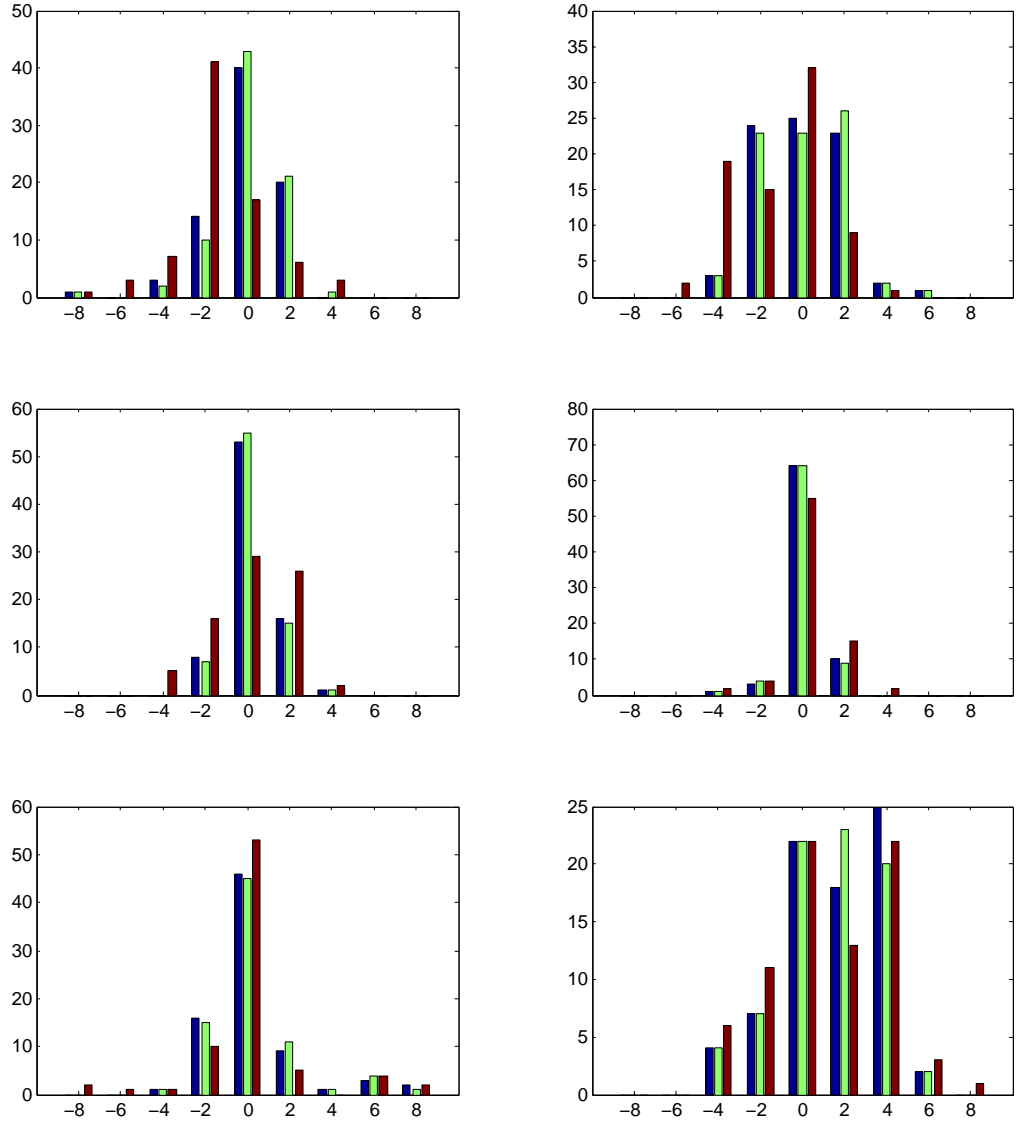}
\caption{{Histograms of the relative errors, in units of bid-ask spread, of the in-sample fit for the VG model (blue bars), EXP model (green bars) and 
Black-Cox model (red bars) for dataset 1 (top), dataset 2 (middle) and dataset 3 (bottom). The tenor on the left is 1-year and on the right, 10-year.}}
\label{histogram}

\end{figure}

 Figure \ref{histogram} displays  histograms of the signed relative error $
(w_t^k)^{-1}{\left( F^k(X_t,\Theta)-Y_t^k\right)}$ for the three models, for the short and long end of the term structure.
For both TCBM models we can see that most errors are bounded by $\pm 2$ and are without obvious bias. By comparison, the errors of the Black-Cox model 
are highly biased downward in the both the short and long terms. For 1-year spreads the majority of errors stay near -2 and for 10-year spreads there is a concentration of errors near -4.  Surprisingly, all the three models perform better and more closely to one another during the crisis period of dataset 2. For the TCBM models, the great majority of errors
are near 0 and without obvious bias. The Black-Cox model does not have obvious bias either, but there are more errors beyond the $\pm 2$ range.
The performance of all three models is better for intermediate tenors between 1 and 10 years, with the mid-range 5-year and 7-year tenors having the best fit. The histograms for these tenors (not shown)
do still indicate that the TCBM models perform better than the Black-Cox model, in regard to both bias and absolute error. 

The estimation results (not shown here) using the Kalman filter method described in Remarks \ref{rm4} are very close to the results shown in
Table \ref{calibration1}, indicating that the transition density can be safely approximated by a gaussian density. The Kalman filter is convenient for calculating the weekly likelihood function, which is needed in the Vuong test \cite{Vuong89},  a test to 
compare the relative performance of nested models.
If $\bar{X}_{t}$ and $\bar P_{t}$ denote the ex-ante forecast and variance of time $t$ values of the measurement series obtained from Kalman filtering, the weekly log-likelihood function can be written as  
\be\label{kalmanlhf}
l_t=-\frac{1}{2}\log|\bar P_{t}|-\frac{1}{2}(\tilde X_{t}-\bar{X}_{t})^\top(\bar P_{t})^{-1}(\tilde X_{t}-\bar{X}_{t})-\sum_k 
f^k(\tilde X_t^k,\Theta).
\ee
The log-likelihood ratio between two models $i$ and $j$ is 
\[
 \lambda_{ij}=\sum_{t=1}^M\left(l_{it}-l_{jt} \right) 
\]
and the Vuong test statistic is
\[
 \CT_{ij}=\frac{\lambda_{ij}}{\hat s_{ij}\sqrt{M}},
\]
where $\hat s_{ij}^2$ is the variance of $\{l_{it}-l_{jt}\}_{t=1,\dots, M}$. Vuong proved that $\CT_{ij}$ is asymptotic to a standard normal under the null hypothesis that
 models $i$ and $j$ are equivalent in terms of likelihood function. Due to
the serial correlation within the 
log-likelihood functions, Newey and West's estimator \cite{NeweWest87} is used for
$\hat s$. The Vuong test results are shown in Table \ref{modelcomp} 
and confirm that 
the Black-Cox model is consistently outperformed by the two TCBM models. Moreover, by this test, the EXP model shows an appreciable improvement over the VG model that could not be easily observed in the previous comparison. 
 
\begin{figure}[b]
\centering\includegraphics[%
  scale=.7]{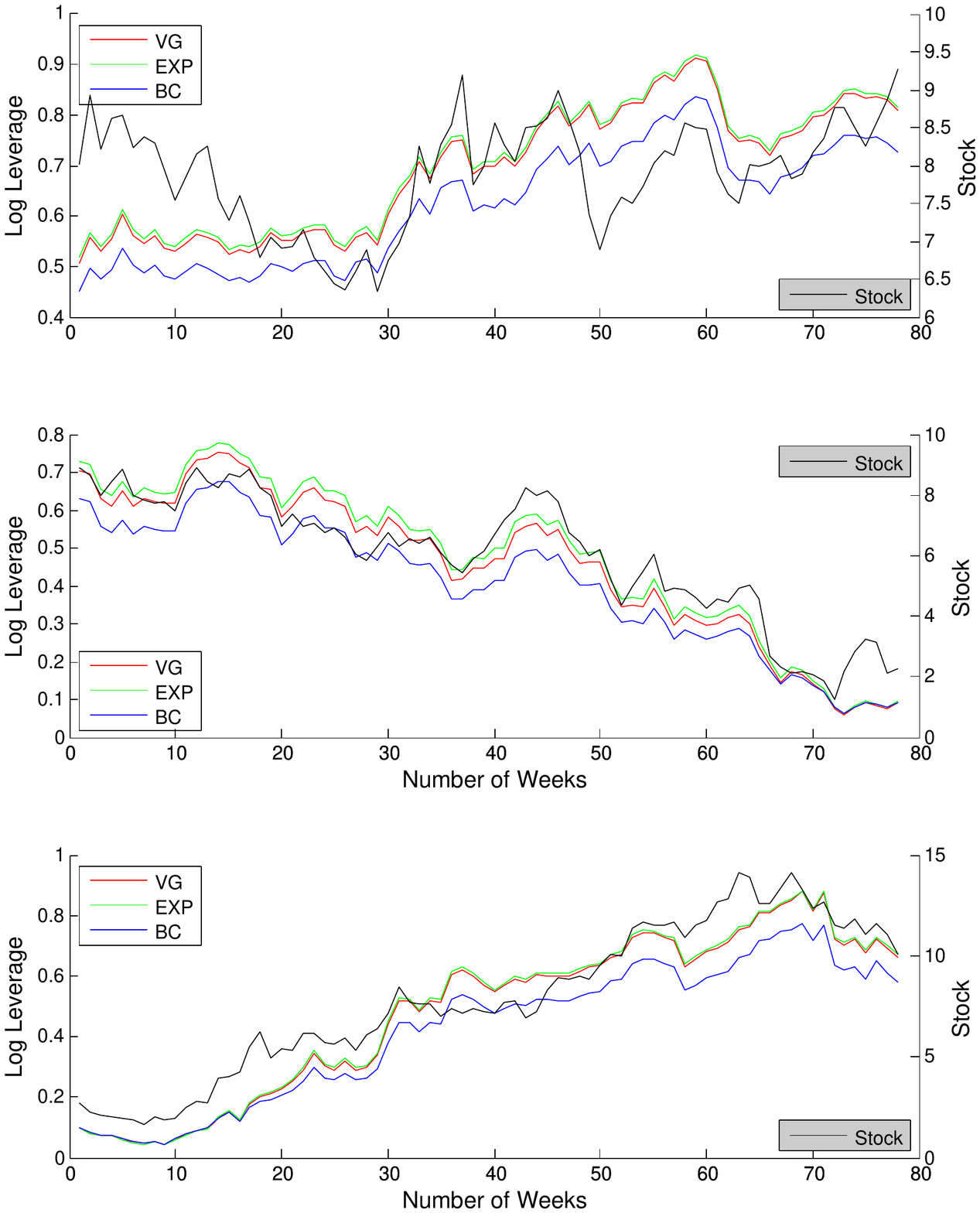}
\caption{{Filtered values of the unobserved log-leverage ratios $X_t$ versus stock price for Ford for dataset 1(top), 2 (middle) and 3 (bottom).  }}
\label{xstockonetwo}
\end{figure}

It is interesting to compare the time series of Ford stock prices to the filtered log-leverage ratios $X_t$. Fig \ref{xstockonetwo} shows there is a strong correlation between these two quantities, indicating that the equity market and credit market are intrinsically connected. The empirical observations supporting this connection and thereafter financial modeling interpreting this connection can be found in \cite{MendCarrLine10}, \cite{CarrWu09} and their references. 

Finally, we mention that a stable model estimation over a 78 week period typically involved about 120 evaluations of the function $\rho(Y,\Theta)$, and took around one minute on a standard laptop.

%\begin{table}[b]
%\centering\begin{tabular}{||c||c||c|c||}
%\hline \hline
%&&Dataset 1&Dataset 2 \tabularnewline
%\hline 
%&number of weeks&78&78\tabularnewline\hline\hline
%&$\hat\sigma$&0.138&0.179\tabularnewline\hline
%VG Model&$\hat{b}$&0.65&0.40\tabularnewline\hline
%&$\hat{c}$&5.72&25.04\tabularnewline\hline 
%&$\hat\beta$&1.63&-7.54\tabularnewline
%\hline\hline
%&$\hat\sigma$&0.150&0.190\tabularnewline\hline
%Exponential Model&$\hat a$&0.022&0.012\tabularnewline\hline 
%&$\hat{b}$&0.58&0.40\tabularnewline
%\hline
%&$\hat\beta$&0.25&-6.39\tabularnewline
%\hline

%
%\end{tabular}

%\label{statmeasure}
%\caption{GMM Calibration results and related statistics for the VG, EXP and SEA models in the statistical measure.}
%\end{table}

%\begin{figure}[b]
%\centering\includegraphics[%
 % scale=.8]{historical3.pdf}
%\caption{{Time series of the 3-year Ford CDS spreads for the three TCBM models, for dataset 1 (top) %%and dataset 2 (bottom).}}
%\label{cdshone1}

%\end{figure}

%\begin{figure}[b]
%\centering\includegraphics[%
 % scale=.8]{historical5.pdf}
%\caption{{Time series of the 5-year Ford CDS spreads for the three TCBM models, for dataset 1 (top) %%and dataset 2 (bottom).}}
%\label{cdshone5}
%\end{figure}

%\begin{figure}[b]
%\centering\includegraphics[%
%  scale=.8]{vggmm.jpg}
%\caption{{VG model density fit in statistical measure for dataset 1(left) and dataset 2(right)}}
%\label{vggmmfit}

%\end{figure}

%\begin{figure}[b]
%\centering\includegraphics[%
%  scale=.8]{levydiffmeas.jpg}
%\caption{{Relative jump measure (EMM vs. statistical) for VG (top) and EXP (bottom) in dataset 1 (left) and dataset 2 (right)}}
%\label{levymeas}

%\end{figure}

\section{Conclusions} 
In this paper, we demonstrated that the Black-Cox first passage model can be efficiently extended to a very broad class of firm 
value processes that includes exponential L\'evy processes. We tested the fit of two realizations of L\'evy subordinated Brownian motion models to observed CDS spreads for Ford Motor Co., a representative firm with an interesting credit history in recent years.  
We found that the 
two L\'evy process models can be implemented very easily, and give similarly good performance in spite of the very different characteristics of  their jump measures. With one extra parameter, both models outperform the Black-Cox model
in fitting the time series of CDS term structures over 1.5 year periods. However, they still have limitations in fitting all tenors of the CDS term structure,
suggesting that further study is needed into models with more flexible time changes.

We also proposed a new method for filtered statistical inference, based on what we call the linearized measurement equation. This new method inductively creates ``quasi-gaussian'' likelihood functions that can be approximated either as truncated gaussians, or as true gaussians in which case we are lead to
a Kalman filter. By their strategic use of the fast Fourier transform, both of our two approximation methods turn out to be very efficient: parameter estimation for a time series of term structures for 78 
weeks can be computed in about a minute. Finally, we observe a strong correlation between Ford's stock price and the filtered values of its unobserved log-leverage ratios. This final observation provides the motivation for our future research that will extend these TCBM credit models to TCBM models for the joint dynamics of credit and equity.

%
%\bigskip
%\noindent{\bf Acknowledgments:\ } 

%\bibliographystyle{abbrvnat}

\bibliographystyle{plain}

\bibliography{truefinancereferences}

\end{document}